# Merging Two Arima Models for Energy Efficiency in WSN

Saumay Pushp

Department of Computer Science and Engineering, KIT, Kanpur,UP-208001, India
saumaypushp@gmail.com

# 1 Abstract

In this report we present the way to merge Arima models in a sensor network and discuss about the error associated with it. The motivation behind merging ARIMA models is to reduce the energy consumption in the sensor nodes. This merging will go in a tree like structure so the overall impact on reducing the energy consumption would be large.

# 2 Introduction

Nowadays Wireless Sensor Networks (WSNs) are deployed in various fields for different operations like controlling nuclear reactors, detecting seismic activities, security and surveillance, navigational activities, industrial automation and others. Such wide-ranging applications requiring WSNs, make them candi dates for intense research. Wireless sensor network consists of a sensor, a radio transceiver or other wireless communication device, a small micro controller and an energy source normally a battery. WSN are deployed in remote areas where recharging of battery might not be possible. So all the operations(sensing and transmitting data) involving battery must be energy efficient, so as to increase the network lifetime. Base station is supposed to be connected with a constant power source and its energy is not considered for calculating network lifetime. Lifetime of a WSN is defined as the time until which the first sensor node runs out of energy. Base station acts as the central object whose task is to collect all the information and process it for further actions. It also helps the network to organize itself. We can reduce power consumption in several ways(1) by using better query processing approach at architecture level (2) by reducing the amount of communication that needs to be done between sensor nodes and its parent node (3) by also reducing the calculation done at the node.For reducing the amount of communication we use ARIMA model.

# 3 Arima Model

Arima model is a stochastic model and stands for Auto regressive Integrated moving average model. It is either used to better understand time series data or to predict the future value of a time series. In Arima model the future value prediction uses some previous data terms of the series. In general arima model is denoted as Arima(p,d,q)
Where :
p is the autoregressive order
d is the non-seasonal difference
q is the order of the moving average component
p,d,q are integers greater than or equal to zero.
For identifying the appropriate arima model we first identify the order(s) of the differencing (d) needed to stationarize the series. After the series is stationarized , it is now basically an ARMA model left which is for the stationary series. Stationary series means having the expected values, variance and auto-covariance constant.[5]

**ARMA model is denoted as ARMA(p,q)**
Where p and q are same as that in ARIMA model.
The auto-regressive part of the model has its origin that individual values of the
time series can be described by linear model based on preceding observations.
The general formula for describing AR[p] (auto-regressive part) is:

$$Y(t) = \sum_{i=1}^{p} \phi_i Y(t-i) \quad [1]$$

Here at time t we are nding x(t) using p previous observations. The order of the model is determined by p. But since time series can receive random shocks in a noisy environment and and may memorize random shocks for a while here moving average part comes into play.The moving average part of the model takes into account the preceding estimation errors. Past estimation or forecasting errors are taken into account when estimating the next time series value.The moving average part captures
the influence of previously received random shocks to the future.The general formula describing Moving average part (MA[q]) is:

$$Y(t) = -\sum_{i=1}^{q} \psi_i e(t-i) \quad [1]$$

Here the di erence between the estimation x(t) and the actually observed value x(t) is denoted by e(t).

When combining both AR and MA models, ARMA models are obtained.The general equation describing ARMA models(ARMA[p,q]) are

$$Y(t) = \sum_{i=1}^{p} \phi_i Y(t-i) - \sum_{i=1}^{q} \psi_i e(t-i) [1]$$

After a suitable ARMA model is tted to the resulting series the estimated forecast is integrated d times to get the predicted value.

## 4 Ways of grouping Arima models in a sensor network

There are several ways in which arima models in a WSN can be merged. In this section we discuss the number of ways of doing so.Consider a wireless sensor network with 2n sensor nodes.Thus we can make n groups of 2 nodes each let number of ways in which 2n nodes can be divided into n groups of 2 nodes each is T(2n):

$$T(2n) = \frac{{}^{2n}C_2}{n} \cdot \frac{{}^{2n-2}C_2}{n-1} \cdot \ldots \cdot \frac{{}^{2}C_2}{1}$$

$$T(2n) = \frac{(2n)!}{(2!)^n \cdot (n!)}$$

$$T(2n) = (2n-1) \cdot T(2n-2)$$

Then these n groups can further be grouped taking 2 groups at a time and thus we will get n=2 such groups. So, going in a similar way we will get a tree of nodes.

let the total number of such tree possible is G(2n)

$$G(2n) = T(2n) \cdot G(n)$$

Consider dividing 2n-1 things into n-1 groups of 2:

$$S = \frac{{}^{2n-1}C_2}{n-1} \cdot \frac{{}^{2n-3}C_2}{n-3} \cdot \ldots \cdot \frac{{}^{5}C_2}{2} \cdot {}^{3}C_2$$

$$S = \frac{(2n-1)!}{(n-1)! \cdot (2!)^{(n-1)}}$$

Thus we can see S = T(2n) i.e. the number of ways 2n-1 things can be divided in a n-1 group of 2 is equal to dividing 2n things in n group of 2.

Thus,
$$G(n) = T(n) \cdot T(n/2) \ldots T(2)$$

whenever n/2 is an odd number we take the next higher even number. Let us take an example:

Now consider a network with 4 nodes :

Using the formula we have:

$$G(4) = T(4) \cdot T(2)$$
$$T(4) = 3 \text{ and } T(2) = 1$$
$$So, G(4) = 3$$

The 3 trees thus formed are:

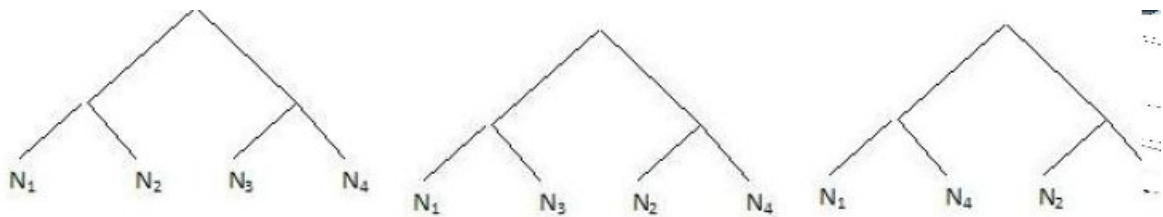

But we do not have to consider all these possibilities in case of sensor network since only the nodes which are close to each other such that they do not have much difference in the values sensed by them are to be considered for merging their Arima models.

**5 Merging two Arima Models :**

In this section we show you how to get an average model given two arima models. Consider the two arima models to be merged are:

$$A_1 : Y_1(t) = \sum_{i=1}^{n} \phi_1^i Y_1(t-i) + \sum_{i=1}^{m} \psi_1^i e_1(t-1)$$
$$A_2 : Y_2(t) = \sum_{i=1}^{n} \phi_2^i Y_2(t-i) + \sum_{i=1}^{m} \psi_2^i e_2(t-1)$$

Average Cofficients are:

$$\phi^i_{avg} = \frac{\phi^i_1 + \phi^i_2}{2}$$

$$\psi^i_{avg} = \frac{\psi^i_1 + \psi^i_2}{2}$$

Apart from this deviation of both the parameters are also stored. So, we keep
$\sigma_\phi, \sigma_\psi$ for both the models.

### 5.1 Error in the average model :

In this section we nd out the error in the average model described above. Let the 2 arima models be:

$$A_1 : Y_1(t) = \phi_1 Y_1(t-1), \text{ with error } \epsilon_1$$
$$A_2 : Y_2(t) = \phi_2 Y_2(t-1), \text{ with error } \epsilon_2$$

So, to next level we send $\phi_1, \phi_2, \epsilon_1, \epsilon_2, Y_1(t-1), Y_2(t-1)$
Here, $Y_1(t-1), Y_2(t-1)$ are sent once and is not sent again till the sensed value lies between some predefined range.
So, $Y(t-1) - \beta \geq Y(t-1) \leq Y(t-1) + \beta$ till then $Y(t-1)$ is not sent.
The average model for the two arima models $A_1, A_2$ are:

$$\phi_{avg} = \frac{\phi_1 + \phi_2}{2}$$
$$Y_{avg}(t) = \phi_{avg} \frac{Y_1(t-1) + Y_2(t-1)}{2}$$

Let the error in the above model be $\epsilon_{avg}$ and $\phi_1 = \phi_2 + d$.
Then:

$$\begin{aligned}
\epsilon_{avg} &= \frac{Y_1(t) + Y_2(t)}{2} - Y_{avg}(t) \\
&= \frac{Y_1(t) + Y_2(t)}{2} - \phi_{avg} \frac{Y_1(t-1) + Y_2(t-1)}{2} \\
&= \frac{\epsilon_1 + \epsilon_2}{2} + \frac{d(Y_1(t-1) - Y_2(t-1))}{4}
\end{aligned}$$

Thus error in the above average model is :

$$\epsilon_{avg} = \frac{\epsilon_1 + \epsilon_2}{2} + \frac{d(Y_1(t-1) - Y_2(t-1))}{4} \tag{10}$$

Now, consider the 2 arima models be:

$$A_1 : Y_1(t) = \sum_{i=1}^{n} \phi_1^i Y_1(t-i) + \sum_{i=1}^{m} \psi_1^i e_1(t-1)$$
$$A_2 : Y_2(t) = \sum_{i=1}^{n} \phi_2^i Y_2(t-i) + \sum_{i=1}^{m} \psi_2^i e_2(t-1)$$

errors in the two models given above are $\epsilon_1$ and $\epsilon_2$.
The average of the two model is:

$$\phi_{avg}^i = \frac{\phi_1^i + \phi_2^i}{2}$$
$$\psi_{avg}^i = \frac{\psi_1^i + \psi_2^i}{2}$$
$$Y_{avg}(t) = \sum_{i=1}^{n} \phi_{avg}^i Y_1(t-i) + \sum_{i=1}^{m} \psi_{avg}^i e_1(t-1)$$

$$error = \frac{\epsilon_1 + \epsilon_2}{2} + \sum_{i=1}^{n} \frac{d_1^i(Y_1(t-i) - Y_2(t-i))}{4} + \sum_{i=1}^{m} \frac{d_2^i(e_1(t-i) - e_2(t-i))}{4} \tag{11}$$

where $d_1^i = \phi_i^1 - \phi_i^2$ and $d_2^i = \psi_i^1 - \psi_i^2$

As we are choosing nodes randomly to take the average so we might end up having 2 arima models which are average of different number of nodes like a model A1 is the average of 10 nodes while a model A2 is the average of just 2 nodes but we are giving equal weight to both models. So, not to encounter such a problem we can send another parameter that gives the number of nodes that model represents. So, instead of taking average we can take in the ratio of the number of nodes in that model.

## 6 Individual models from the average model :

In this section we assume that we have the average arima model and we want to find the two individual models, this average consists of. For this we have the average of each individual parameters and we have the deviation of the parameter so we can find out the approximate parameters.

Let we have:

$\phi_{avg}^1, \phi_{avg}^2, \phi_{avg}^3$ which are the average parameters of nodes S1 and S2 and deviation of $\phi, \sigma_1, \sigma_2$ for both the nodes.

Then we can say that for node S1:

$$\phi_1 = \phi_{avg}^1 \pm \sigma_1$$
$$\phi_2 = \phi_{avg}^2 \pm \sigma_1$$
$$\phi_3 = \phi_{avg}^3 \pm \sigma_1$$

Similarly for node 2 (S2).

## 7 Example :

**In this I will show you an example of a 16 node sensor network and apply the average model till we are left with a single Arima model.**
**The data from the nodes are given in the tables below:**

| Node1 | Node2 | Node3 | Node4 | Node5 | Node6 | Node7 | Node8 |
|---|---|---|---|---|---|---|---|
| 94.5267 | 91.5259 | 102.8402 | 103.6 | 112.983 | 114.757 | 119.7618 | 123.5316 |
| 95.2966 | 91.3538 | 103.62 | 104.398 | 112.846 | 115.56 | 120.7324 | 123.4532 |
| 94.4962 | 90.9694 | 104.53 | 104.2 | 112.6898 | 115.16 | 121.6346 | 122.578 |
| 95.3877 | 91.6326 | 104.195 | 104.9686 | 112.56 | 115.0513 | 122.4855 | 123.5098 |
| 95.1044 | 92.5228 | 103.917 | 105.52 | 113.559 | 114.833 | 121.9523 | 124.4406 |
| 94.752 | 92.174 | 103.44 | 106.15 | 114.072 | 115.672 | 122.492 | 125.1615 |
| 95.5597 | 92.1098 | 103.0753 | 106.665 | 114.6846 | 116.309 | 121.7412 | 124.5929 |
| 96.4787 | 92.0698 | 104.027 | 107.581 | 114.3886 | 116.834 | 122.5014 | 125.3314 |
| 96.3392 | 91.1544 | 104.66 | 108.298 | 113.895 | 117.806 | 123.0139 | 125.9713 |
| 97.2885 | 91.0282 | 104.52 | 108.9 | 113.602 | 117.035 | 123.6816 | 125.2631 |

| Node9 | Node10 | Node11 | Node12 | Node13 | Node14 | Node15 | Node16 |
|---|---|---|---|---|---|---|---|
| 131.035 | 135.786 | 141.6878 | 144.9564 | 151.7938 | 154.895 | 159.7049 | 161.813 |
| 130.6034 | 135.1712 | 141.356 | 145.545 | 151.5418 | 155.4165 | 159.4737 | 161.5432 |
| 131.5536 | 134.7242 | 140.4758 | 146.2023 | 152.0372 | 155.24 | 160.0581 | 162.0634 |
| 132.4737 | 134.4981 | 141.3558 | 147.0611 | 152.7976 | 154.7603 | 160.3025 | 161.907 |
| 132.1785 | 134.8731 | 142.185 | 146.1821 | 153.7823 | 155.558 | 160.15 | 162.7047 |
| 133.0595 | 135.4255 | 141.5244 | 147.106 | 154.7173 | 156.29 | 160.8823 | 162.5 |
| 133.7 | 135.9819 | 141.0666 | 146.3092 | 154.0329 | 156.9471 | 160.7568 | 163.4675 |
| 132.8368 | 136.8149 | 141.9599 | 147.1239 | 154.416 | 156.0189 | 161.5502 | 164.2871 |
| 133.4563 | 136.4757 | 141.6096 | 147.8081 | 155.1657 | 156.6583 | 161.3861 | 164.1381 |
| 132.8939 | 137.3825 | 142.2962 | 148.719 | 154.7971 | 157.418 | 162.1312 | 165.088 |

Here we can see the maximum difference between the data from the nodes is around 75% . Next we will build Arima models each of the nodes. The parameters for all the arima models are (p,d,q)=(3,0,0).

The arima models build using the data from the table above are given below:

| Constant | Ar1 | Ar2 | Ar3 | Error Value |
|---|---|---|---|---|
| 91.7057 | 0.6274 | -0.3723 | -0.1651 | 0.487 |
| 95.6915 | 0.5471 | 0.4443 | -0.2682 | 0.8282 |
| 103.8353 | 0.3643 | -0.5858 | -0.4035 | 0.3504 |
| 106.6934 | 1.0017 | 0.763 | -0.8633 | 0.3726 |
| 113.5179 | 0.9036 | -0.0513 | -0.4407 | 0.4002 |
| 115.7786 | 0.8245 | 0.2181 | -0.4696 | 0.7188 |
| 121.8048 | 1.0239 | 0.2242 | -0.4548 | 0.8045 |
| 124.3562 | 0.8766 | -0.2091 | -0.013 | 0.8523 |
| 132.2516 | 0.7294 | 0.0796 | -0.1066 | 0.8585 |
| 135.9154 | 0.7582 | 0.5832 | -0.8245 | 0.4165 |
| 141.5312 | 0.0085 | -0.3416 | 0.1087 | 0.6117 |
| 146.795 | 0.8617 | 0.3721 | -0.4524 | 0.8828 |
| 153.4567 | 1.1943 | -0.3886 | -0.0531 | 0.7704 |
| 155.9707 | 0.8218 | -0.3058 | 0.2852 | 0.7818 |
| 160.7683 | 0.8098 | 0.7403 | -0.6838 | 0.4343 |
| 163.2291 | 0.8607 | 0.5865 | -0.6109 | 0.6274 |

Now we will apply the average model on Arima model of node 1 and arima model of node 2 and so on.

The corresponding Arima models thus obtained are given below:

| Constant | Ar1 | Ar2 | Ar3 | Error Value | Error(%) |
|---|---|---|---|---|---|
| 93.6986 | 0.5872 | 0.036 | -0.2167 | 2.4219 | 2.64 |
| 105.2643 | 0.683 | 0.0886 | -0.6334 | 2.2034 | 2.122 |
| 114.6482 | 0.8641 | 0.0834 | 0.4551 | 1.5463 | 1.362 |
| 123.0805 | 0.9502 | 0.0076 | -0.2339 | 2.2604 | 1.856 |
| 134.0835 | 0.7438 | 0.3314 | -0.4655 | 3.1735 | 2.4 |
| 144.1631 | 0.4351 | 0.0152 | -0.1718 | 3.8556 | 2.724 |
| 154.7137 | 1.008 | -0.3472 | 0.116 | 2.4349 | 1.587 |
| 161.9987 | 0.8352 | 0.6634 | -0.6473 | 1.9558 | 1.22 |

As we can see the error is around 2-3% so we can apply this model for this level Further we will apply the average model on Arima models obtained above to get the models at the grand-father level.

The models thus obtained are:

| Constant | Ar1 | Ar2 | Ar3 | Error Value | Error(%) |
|---|---|---|---|---|---|
| 99.4815 | 0.6351 | 0.0623 | -0.4251 | 9.8466 | 10.5 |
| 118.8644 | 0.9071 | 0.0455 | 0.1106 | 8.056 | 7.03 |
| 139.1233 | 0.5895 | 0.1733 | -0.3187 | 10.836 | 8.896 |
| 158.3562 | 0.9306 | 0.1581 | -0.2657 | 9.332 | 6.0812 |

In this level the maximum error is around 10%. Lets the apply the average model further.

| Constant | Ar1 | Ar2 | Ar3 | Error Value | Error(%) |
|---|---|---|---|---|---|
| 109.1729 | 0.7711 | 0.0539 | -0.1573 | 22.6679 | 24.72 |
| 148.7398 | 0.76 | 0.1657 | -0.2922 | 24.28 | 18.267 |

Here the error is around 25% but then this is the average model for 8 nodes which have their data ranging from 94 to 126.

Now lets get the arima model for the base station.

| Constant | Ar1 | Ar2 | Ar3 | Error Value | Error(%) |
|---|---|---|---|---|---|
| 128.9563 | 0.7656 | 0.1098 | -0.2248 | 48.9589 | 53.38 |

Here the value of error is around 49 and taken on the minimum value of node data the percentage error is 53.38%. So we can see that the model till the third level does not have much error.